\documentclass{PoS}

\usepackage{amssymb}
\usepackage{amsmath}

\def\beq{\begin{equation*}}
\def\eeq{\end{equation*}}
\newcommand{\bc}{\begin{center}}
\newcommand{\ec}{\end{center}}
\newcommand{\bea}{\begin{eqnarray}}
\newcommand{\eea}{\end{eqnarray}}
\newcommand{\bean}{\begin{eqnarray*}}
\newcommand{\eean}{\end{eqnarray*}}
\newcommand{\vev}[1]{\langle #1 \rangle }

\newcommand{\eps}{\epsilon}
\newcommand{\ep}{\epsilon}
\newcommand{\ord}{{\cal O}}

\def\a{\alpha}

\def\a{{\alpha}}

\def\b{{\beta}}

\def\c{{\gamma}}
\def\g{{\gamma}}

\def\half{\frac{1}{2}}

\newcommand{\hypgeo}[4]{\,_2F_1\left(#1,#2;#3;#4\right)}

\def\Li{{\rm Li}}

\renewcommand{\ln}{\log}

\newcommand{\hb}[1]{#1}

\def\bsp#1\esp{\begin{split}#1\end{split}}
\DeclareMathOperator*{\res}{Res}

\title{Generalized hypergeometric functions and intersection theory for Feynman integrals}

\ShortTitle{Hypergeometric coaction}

\author{Samuel Abreu\\
		Center for Cosmology, Particle Physics and
		Phenomenology (CP3),
		Universit\'e Catholique de
		Louvain, 1348 Louvain-La-Neuve, Belgium\\
        E-mail: \email{samuel.abreu@uclouvain.be}}

\author{\speaker{Ruth Britto}
		\\
        School of Mathematics, Trinity College, Dublin 2,
        Ireland ;\\
        Hamilton Mathematics Institute, Trinity
        College, Dublin 2, Ireland ;\\
		Institut de Physique Th\'eorique,
		Universit\'e Paris Saclay,
		CEA, CNRS, F-91191 Gif-sur-Yvette
		cedex, France\\
        E-mail: \email{britto@maths.tcd.ie}}
        
\author{Claude Duhr\\
        Theoretical Physics Department, CERN, Geneva,
        Switzerland ;\\
        E-mail: \email{claude.duhr@cern.ch}}

\author{Einan Gardi\\
        Higgs Centre for Theoretical Physics, School of
        Physics and Astronomy, The University of Edinburgh,
        Edinburgh EH9 3FD, Scotland, UK\\
        E-mail: \email{einan.gardi@ed.ac.uk}}

\author{James Matthew\\
        Higgs Centre for Theoretical Physics, School of
        Physics and Astronomy, The University of Edinburgh,
        Edinburgh EH9 3FD, Scotland, UK\\
        E-mail: \email{james.matthew@ed.ac.uk}}

\abstract{Feynman integrals that have been evaluated in dimensional regularization can be written in terms of generalized hypergeometric functions. It is well known that properties of these functions are revealed in the framework of intersection theory.
We propose a new application of intersection theory to construct a coaction on generalized hypergeometric functions. When applied to dimensionally regularized Feynman integrals, this coaction reproduces the coaction on multiple polylogarithms order by order in the parameter of dimensional regularization.
 }

\FullConference{14th International Symposium on Radiative Corrections (RADCOR2019)\\ 
9-13 September 2019\\
		Palais des Papes, Avignon, France}

\ifx \tadInsert \undefined
	\def\tadInsert [#1]{
		\raisebox{-4mm}{\includegraphics[keepaspectratio=true, width=.8cm]{./diagrams/#1}}
	}
\fi

\ifx \tadInsertLow \undefined
	\def\tadInsertLow [#1]{
		\raisebox{-4.1mm}{\includegraphics[keepaspectratio=true, width=.81cm]{./diagrams/#1}}
	}
\fi

\ifx \bubInsertHigh \undefined
	\def\bubInsertHigh [#1]{
		\raisebox{-2.2mm}{\includegraphics[keepaspectratio=true, width=2cm]{./diagrams/#1}}
	}
\fi

\ifx \bubInsert \undefined
	\def\bubInsert [#1]{
		\raisebox{-3.6mm}{\includegraphics[keepaspectratio=true, width=2cm]{./diagrams/#1}}
	}
\fi

\ifx \bubInsertLow \undefined
	\def\bubInsertLow [#1]{
		\raisebox{-4.2mm}{\includegraphics[keepaspectratio=true, width=2cm]{./diagrams/#1}}
	}
\fi

\ifx \triInsert \undefined
	\def\triInsert [#1]{
		\raisebox{-6.3mm}{\includegraphics[keepaspectratio=true, width=2cm]{./diagrams/#1}}
	}
\fi

\ifx \triInsertLow \undefined
	\def\triInsertLow [#1]{
		\raisebox{-6.9mm}{\includegraphics[keepaspectratio=true, width=2cm]{./diagrams/#1}}
	}
\fi

\ifx \boxInsert \undefined
	\def\boxInsert [#1]{
		\raisebox{-4.9mm}{\includegraphics[keepaspectratio=true, width=2.3cm]{./diagrams/#1}}
	}
\fi

\ifx \boxInsertLow \undefined
	\def\boxInsertLow [#1]{
		\raisebox{-7.45mm}{\includegraphics[keepaspectratio=true, width=2.3cm]{./diagrams/#1}}
	}
\fi

\begin{document}

\section{Introduction}

The algebraic structure known as {\em coaction} has recently played an interesting role in computing Feynman diagrams and understanding their analytic structure. In dimensional regularization with $D=n-2\eps$ dimensions where $n$ is even, large classes of loop integrals including all one-loop integrals are expressible in terms of multiple polylogarithms (MPLs) when expanded in $\eps$. MPLs are equipped with a coaction
\cite{2001math......3059G,2002math......8144G,B:MTMZ,Duhr:2012fh},
a mathematical operation that exposes properties of MPLs through a decomposition into simpler functions. In particular, since the coaction is naturally compatible with the actions of differential operators and taking discontinuities across branch cuts, it can be considered as a computational tool.

There is a natural mathematical coaction on Feynman integrals~\cite{Brown:coaction, Panzer:2016snt,Brown:2015fyf}. The coaction captures information about discontinuities, which for Feynman integrals can be expressed in terms of unitarity cuts and their generalizations. 
Motivated by these observations,
it was conjectured that there exists a diagrammatic coaction corresponding precisely to the well known coaction on MPLs,
for the class of Feynman graphs admitting an expansion in MPLs. 
 This conjecture was expressed 
in \cite{Abreu:2017enx,Abreu:2017mtm}, which further proposed a general form for a coaction on integrals:
\begin{equation}\label{eq:masterformula}
	\Delta\left(\int_\gamma\omega\right)=
	\sum_{ij}c_{ij}\int_\gamma\omega_i\otimes
	\int_{\gamma_j}\omega\,.
\end{equation}
Here the $\omega_i$ are generators of a suitably defined cohomology group of which $\omega$ is also an element, and the $\gamma_j$ are likewise generators of a suitable homology group of which $\gamma$ is also an element. 
It is straightforward to check that eq. (\ref{eq:masterformula}) satisfies the algebraic properties of a coaction, such as  coassociativity.

In the case of Feynman integrals, the cohomology group contains the usual integrands constructed from Feynman rules. In turn, the homology group contains the original contour in addition to contours encircling poles giving rise to the residues associated to generalized unitarity cuts. The coefficients $c_{ij}$ remain to be determined. In the case of one-loop Feynman integrals, a basis has been found in which the $c_{ij}$ are known precisely \cite{Abreu:2017enx,Abreu:2017mtm}.\footnote{For extensions to two loops, see the contribution of James Matthew to these proceedings.}

Consider for example the following simple scalar 1-loop integral, evaluated in $4-2\eps$ dimensions, where the thick lines represent a propagator of mass $m$ and a non-null external leg of momentum $p$, while the thin lines represent massless propagators and external legs carrying null momenta:
\begin{align}\label{eq:t1m1exp}
	&\triInsertLow[t1m1LabelEdges]
	=\frac{1}{p^2}\left[\frac{\ln\left(\frac{m^2}{m^2-p^2}\right)}{\epsilon}+\text{Li}_2\left(\frac{p^2}{m^2}\right)+\log ^2\left(1-\frac{p^2}{m^2}\right)+\log (m^2) \log\left(1-\frac{p^2}{m^2}\right)\right]+\mathcal{O}\left(\epsilon\right)\,.
\end{align}
The coaction of the Laurent expansion on right-hand side can be evaluated directly, to arbitrary orders in $\eps$. The terms written explicitly can be treated with well-known formulas for the coaction of polylogarithms, such as
\bean
\Delta(\ln z) &=& 1 \otimes \ln z + \ln z \otimes 1\,, \\ 
\Delta(\ln^2 z) &=& 1 \otimes \ln^2 z +  2 \ln z \otimes \ln z + \ln^2 z \otimes 1\,, \\
\Delta(\Li_2(z)) &=& 1 \otimes \Li_2(z) + \Li_2(z) \otimes 1  + \Li_{1}(z) \otimes \ln z\,.
\eean
We compare the result of that coaction with the diagrammatic coaction on the left-hand side, which was claimed in \cite{Abreu:2017enx,Abreu:2017mtm} to be
\begin{align}
\label{eq:t1m1diagcoac}
	\Delta\left[\triInsertLow[t1m1LabelEdges]\right]=
	\tadInsert[tad1]\otimes \left( \triInsertLow[t1m1Cut1LabelEdges] + \half \triInsertLow[t1m1Cut12LabelEdges]\right)
	+
	\bubInsertLow[bub1m1LabelEdges] \otimes\triInsertLow[t1m1Cut12LabelEdges]\,.
\end{align}
The claim is that when  each of these graphs is evaluated according to Feynman rules for scalar fields, with specific prescriptions for cut lines and dimensionality, then, order by order in $\eps$, the result will agree with the coaction on MPLs. 

We now imagine summing the full Laurent series. 
The example shown above has a closed-form expression in terms of Gauss's hypergeometric function,
\begin{equation}\label{eq:t1m1hyp}
	\triInsertLow[t1m1LabelEdges]
	=\frac{e^{\gamma_E\epsilon}\Gamma(1+\epsilon)}{\epsilon(1-\epsilon)}(m^2)^{-1-\epsilon}\hypgeo{1}{1+\epsilon}{2-\epsilon}{\frac{p^2}{m^2}},
	\end{equation}
	and each of the integrals on the right-hand side of eq.\ (\ref{eq:t1m1diagcoac}) is likewise a ${}_2F_1$ function, or a simpler function arising from a special configuration of the first three arguments.
Many other Feynman integrals in dimensional regularization are known to evaluate to hypergeometric-type functions whose Laurent expansions in $\eps$ are given by MPLs.
We now conjecture a coaction on generalized hypergeometric functions of this type that is compatible with the coaction on Feynman integrals, as well as  with the coaction on MPLs order by order in  the $\eps$-expansion.

There are various ways to define generalized hypergeometric functions. Here we define them as integrals and use the framework of twisted (co)homology \cite{AomotoKita}. We propose a coaction of the form (\ref{eq:masterformula}) in which the coefficients $c_{ij}$ can  be derived from so-called intersection numbers. 
In this article, we state our conjecture more explicitly and illustrate it with a couple of examples.

This contribution to the proceedings of RADCOR2019 is based on work that has since appeared in \cite{Abreu:2019wzk}, which contains many more details and examples, and a fuller discussion.

\section{Generalized hypergeometric functions and intersection numbers}

We consider integrals of the form $\int_\gamma \omega$, where $\omega$ is a cohomology class represented by the differential $n$-form 
\begin{equation}\label{eq:factorizedIntegrand}
	\omega=d\mathbf{u}\prod_{I} P_I(\mathbf{u})^{\alpha_I}\,,
\end{equation}
where $d\mathbf{u}=du_1\wedge\ldots\wedge du_n$,
 the $P_I$ are irreducible polynomials in the variables $u_i$, 
and  $\alpha_I\in\mathbb{C}$. 
We further assume that the exponents take the form $\alpha_I=n_I+a_I\epsilon$, where $n_I\in\mathbb{Z}$,
$a_I\epsilon\in\mathbb{C}^*$, $\sum_I a_I \neq 0$, and where $\epsilon$ can be taken to be infinitesimally small.
We define the decomposition
$\omega=\Phi\varphi$ where
\begin{equation}
\label{eq:Phiphi}
\Phi = \prod_{I} P_I({\bf u})^{a_I \eps}\quad \textrm{and} \quad
\varphi = d{\bf u} \prod_{I} P_I({\bf u})^{n_I}\,.
\end{equation}
Here $\Phi$ is a multivalued function, while $\varphi$ is a single-valued differential form.
The integration contour $\gamma$ is chosen to have its boundary contained within the algebraic variety $\prod_I P_I (\mathbf{u})=0$.

The natural mathematical framework to discuss such integrals, which we identify as generalized hypergeometric functions, is that
of twisted homology and cohomology 
\cite{AomotoKita}.\footnote{See also 
refs.~\cite{Mizera:2017rqa,Mastrolia:2018uzb,
Frellesvig:2019uqt,Frellesvig:2019kgj}, which are featured in the contributions of Hjalte Frellesvig and Manoj Mandal to these proceedings.}
In our notation introduced above, $\omega$ is an element of a twisted cohomology group, while $\gamma$ is an element of a twisted homology group. The {\em twist} is the single-valued function $d\log\Phi$, and it is the multi-valuedness of $\Phi$ that gives rise to the twisted structure.
The mathematical literature contains a great deal of information about these types of twisted   homology and cohomology groups. We would like to identify bases $\{\gamma_j\}$ and 
$\{\omega_i\}$ of these groups in order to construct a coaction of the form (\ref{eq:masterformula}). In general, this is a difficult problem. 
We restrict our attention to the case of so-called positive geometries \cite{Arkani-Hamed:2017tmz},
which are sufficient for the known examples of Feynman diagrams with MPL expansions. For these functions, it is possible to identify a basis of homology classes, represented by integration contours, for which one can construct corresponding {\em canonical} differential forms having logarithmic singularities and unit residues precisely at the boundary of the contour.
In our examples where the $P_k({\bf u})$ are mostly linear, there is a natural choice of the basis of integration contours $\{\gamma_j\}$, and it is straightforward to construct their canonical forms. In cases where the canonical forms can be written in the form $\bigwedge_i d\log y_i({\bf u})$, we refer to them as dlog forms.

Stokes' theorem implies that 
$	\int_\gamma\Phi\varphi=
	\int_\gamma\Phi(\varphi+\nabla_\Phi\xi)$ for an arbitrary smooth $(n-1)$-form $\xi$, so $\varphi$ is a \hb{twisted} cohomology class.
The {\em intersection number} pairing  of twisted cohomology classes (forms) is defined by
\bea
	\langle \varphi_i,\psi_j\rangle_\Phi
	=\frac{1}{(2\pi i)^2}\int
	\iota_\Phi(\varphi_i)\wedge\psi_j\,,
\eea
where $\iota_\Phi$ denotes selection of a representative of the cohomology class with compact support.
To compute intersection numbers in practice, it is preferable to use equivalent formulas based on residues \cite{Mizera:2017rqa,Mastrolia:2018uzb}. 
For the case of dlog forms in one variable, we have 
\bea
	\vev{\varphi_i,\psi_j}_\Phi
	=\sum_{u_p}
	\frac{\res_{u=u_p}\varphi_i\,\res_{u=u_p}\psi_j}
	{\res_{u=u_p}d\ln\Phi}\,,
\eea
where the  $u_p$ are the poles of $d\ln\Phi$. 

To construct a coaction formula for a certain hypergeometric function with an integral representation $\int_\gamma \omega$, we use the following procedure.
\begin{itemize}
\item Check that $\omega$ can be written in the form of eq. (\ref{eq:factorizedIntegrand}) subject to the conditions expressed below the formula, and that the boundary of $\gamma$ is contained within the algebraic variety $\prod_I P_I (\mathbf{u})=0$. 
\item Compute the dimensions of the nontrivial cohomology and homology groups, which are equal to each other.  Let us call this number $r$. One method to determine it is to compute the number of critical points of the function $\Phi$, i.e. the number of solutions to the equation $d\log\Phi=0$. This number is an upper bound for $r$, which happens to be saturated for large classes of hypergeometric functions, including all of the examples we have studied.
\item Identify a basis of cohomology $\{\varphi_i\}$ by selecting a set of  $r$ $n$-forms and computing the matrix of their intersection numbers. If the matrix has full rank, then the differential forms are linearly independent and can be taken as a basis.
\item Identify a basis of homology $\{\gamma_j\}$ by selecting a set of $r$ integration contours bounded by some of the algebraic varieties $P_I (\mathbf{u})=0$ and possibly also extending to infinity.\footnote{Hyperplanes at infinity can be included by going to projective space and then imposing the condition that the sum of all $\alpha_I$ be $0$.} We assume that we can choose these contours to be positive geometries. Therefore each $\gamma_j$ has an associated canonical form $\Omega(\gamma_j)$. Linear independence of the homology classes can be checked by computing the matrix of the intersection numbers $\vev{\Omega(\gamma_i),\Omega(\gamma_j)}_\Phi$ and verifying that it has rank $r$.
\item The intersection matrices in the previous steps were computed only to check linear independence of the proposed bases. We now use both bases to compute the intersection matrix needed to construct the coaction:
\bea \label{eq:intmatrix}
\vev{\varphi_i,\Omega(\gamma_j)}_\Phi
\eea
\item 
The elements of the {\em inverse} of the intersection matrix (\ref{eq:intmatrix}) are the coefficients $c_{ij}$ in the coaction formula (\ref{eq:masterformula}) with the integrands $\omega_i = \Phi\varphi_i$ and the contours $\gamma_j$ chosen above.
\end{itemize}
In this procedure, we have constructed two sets of differential forms: one consisting of the canonical forms of the basis of integration contours, and a second set chosen as a basis of integrands in twisted cohomology.
The canonical forms of contours could also be taken directly as the basis of integrands. However, the examples that follow will show that making independent choices of the two bases can lead to intersection matrices that are more nearly diagonal, and thus to a coaction formula with fewer terms, if desired.

\section{Coaction of ${}_2F_1$}

The simplest nontrivial hypergeometric function is Gauss's ${}_2F_1$ function.
This function admits an Euler-type integral representation of the form
\begin{equation}
\label{2F1}
 {}_2F_1(\a,\b;\c;x)=\frac{\Gamma(\c)}{\Gamma(\a)\Gamma(\c-\a)}
 \int_0^1du\,u^{\a-1}(1-u)^{\c-\a-1}(1-ux)^{-\b}\,,
\end{equation}
provided that the integral converges. As discussed above, we restrict our attention to the case where  $\a,\b,\c$ have the form $n_i+a_i\eps$, with
$n_i \in\mathbb{Z}$. Under these conditions, the Laurent expansion is given in terms of MPLs.

To construct the coaction, we first disregard the gamma-function prefactors and study the family of integrals of the form
\bean
  \hb {\int_\gamma \omega} &\equiv&
  \int_0^1du\, u^{n_0+ a_0\eps} (1-u)^{n_1+a_1\eps} (1-ux)^{n_{x}+a_{x}\eps}\,,
  \\
	\omega & = &  \hb{\Phi \varphi}\,, \\
	\Phi &=& u^{a_0\eps} (1-u)^{a_1\eps} (1-ux)^{a_{x}\eps}\,, \\
	\varphi &=& du\,u^{n_0} (1-u)^{n_1} (1-ux)^{n_{x}}\,.
\eean
Here the algebraic variety $\prod_I P_I (\mathbf{u})=0$ is the set of finite branch points, $0, 1$, and $1/x$.
We see that $\varphi$ is a single-valued differential form, and 
$\Phi$ is a multi-valued function,
from which we construct the \hb{twist} 1-form  \hb{$d\log\Phi$}:
\bean
d\log\Phi &=& a_0 \frac{du}{u} -a_1 \frac{du}{1-u}-xa_x \frac{du}{1-ux}\,.
\eean
The equation $d\log\Phi=0$ has two solutions, so the upper bound on the dimensionality of the nontrivial (co)homology groups is 2. In this case it is easy to see that the bound is saturated. The fact that the space of differential forms related through integer shifts of the exponents is 2-dimensional can be understood as expressing Gauss's contiguity relations. 
One can also visibly identify sensible integration contours connecting pairs of the finite branch points, and see that only two of them can be linearly independent.
Consider two such independent contours with boundaries at the branch points, for example
$\gamma_1=[0,1],~~\gamma_2=[0,1/x]$.
From these contours, construct their associated canonical forms \cite{Matsumoto1998,Arkani-Hamed:2017tmz}
\begin{equation*}
	\Omega(\gamma_1)=d\log\frac{u-1}{u}\,,
	\qquad
	\Omega(\gamma_2)=d\log\frac{u-1/x}{u}\,.
\end{equation*}

Since it is possible to use these same forms as the basis of twisted cohomology, let us first simply set
$\widetilde\varphi_1=\Omega(\gamma_1), ~\widetilde\varphi_2=\Omega(\gamma_2)$.
Then the entries of the intersection matrix are
\bean
\begin{array}{ll}	
	\vev{\widetilde\varphi_1,\Omega(\gamma_1)}_\Phi=\dfrac{1}{a_0\epsilon}+\dfrac{1}{a_1\epsilon}\,,  
& \qquad \vev{\widetilde\varphi_1,\Omega(\gamma_2)}_\Phi=\dfrac{1}{a_0\epsilon}\,,
	\\[3mm]
	\vev{\widetilde\varphi_2,\Omega(\gamma_1)}_\Phi=\dfrac{1}{a_0\epsilon}\,,
& \qquad \vev{\widetilde\varphi_2,\Omega(\gamma_2)}_\Phi=\dfrac{1}{a_0\epsilon}+\dfrac{1}{a_{x}\epsilon}\,.
\end{array}
\eean
The matrix can be inverted to produce a coaction formula.

However, a different choice of basis of differential forms leads to a cleaner coaction formula. Suppose that instead we choose $\varphi_1$ and $\varphi_2$ such that their dlog singularities overlap only with the upper boundaries of $\gamma_1$ and $\gamma_2$ respectively, leading to a diagonal intersection matrix. Moreover, let us normalize the differential forms so that the intersection matrix is in fact the identity. Concretely, we take
\begin{equation}\label{eq:froms2f1}
\varphi_1 = -a_1\epsilon d\log (1-u) = a_1\epsilon \frac{du}{1-u}\,, \qquad 
\varphi_2 = - a_{x}\epsilon d\log(1-xu)=a_{x}\epsilon x\frac{du}{1-xu}\,.
\end{equation}
Then the coaction can be written simply as 
\begin{equation}\label{eq:2F1_coac_diag_basis}
	\Delta_\epsilon \int_{\gamma_k}\Phi\varphi_l
	=\int_{\gamma_k}\Phi\varphi_1\otimes\int_{\gamma_1}\Phi\varphi_l
	+\int_{\gamma_k}\Phi\varphi_2\otimes \int_{\gamma_2}\Phi\varphi_l\,.
\end{equation}
We have checked this coaction formula through order $\epsilon^4$, by verifying
that order by order in $\epsilon$ we reproduce the coaction  on MPLs.

If desired, the gamma-function prefactors in eq. (\ref{2F1}) can be restored using the identity
\bean
\Delta\left(e^{\gamma_E\eps}\,\Gamma(1+\eps)\right) &\,= e^{\gamma_E\eps}\,\Gamma(1+\eps)\otimes e^{\gamma_E\eps}\,\Gamma(1+\eps)\,,
\eean
from which it follows that 
\bean
\Delta\left(e^{a\gamma_E\eps}\,\Gamma(m+a\eps)\right) &\,= e^{a\gamma_E\eps}\,\Gamma(1+a\eps)\otimes e^{a\gamma_E\eps}\,\Gamma(m+a\eps)\,,
\eean
for integer values of $m$. The exponential factors will cancel between the numerator and the denominator of eq.~(\ref{2F1}).
We thus obtain a coaction on ${}_2F_1$, in which each entry in turn can be expressed in terms of ${}_2F_1$ functions. The coaction is given explicitly by
\begin{align}
\nonumber\Delta\Big({}_2F_1(\a,\b;\g;x)\Big) &=
{}_2F_1(1+a\eps,b\ep;1+c\ep;x) \otimes {}_2F_1(\a,\b;\g;x) \\
\label{eq:coaction2F1}&- \frac{b\ep}{1+c\ep}\,
{}_2F_1(1+a\ep,1+b\ep;2+c\ep;x) \\
\nonumber& ~~\otimes \frac{\Gamma(1-\b)\Gamma(\g)}{\Gamma(1-\b+\a)\Gamma(\g-\a)}
x^{1-\a}{}_2F_1\left(\a,1+\a-\g;1-\b+\a;\frac{1}{x}\right)\,,
\end{align}
where $\alpha=n_\alpha+a\eps$, $\b=n_\b+b\eps$ and
$\g=n_\g+c\eps$.

\section{Coaction of Appell $F_3$}

\begin{figure}
\begin{center}
\includegraphics[width=7cm]{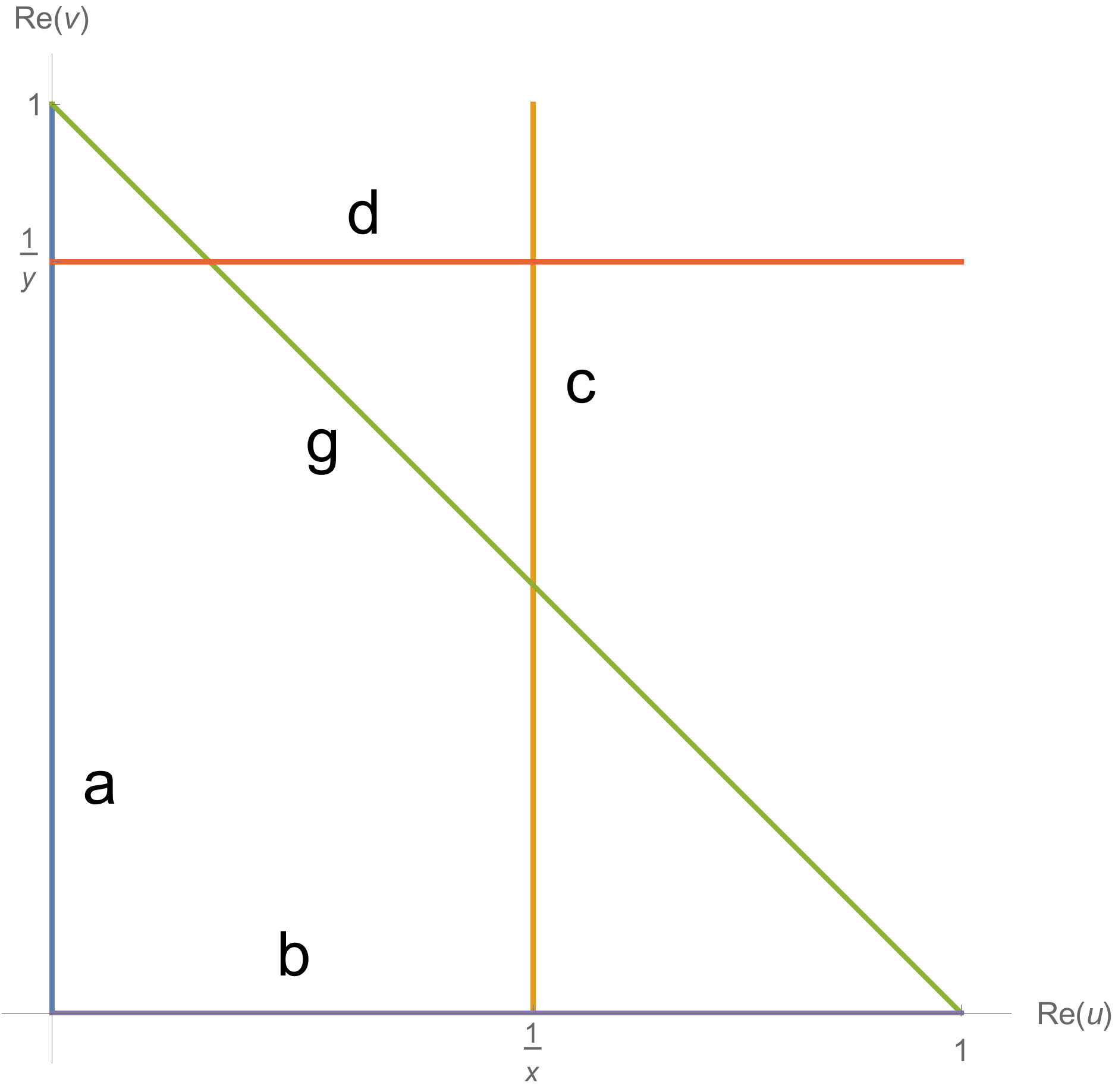}
\end{center}
\caption{Appell $F_3$ is related to an arrangement of hyperplanes.}
\label{fig:f3}
\end{figure}

We now consider  the Appell $F_3$ function, which
has the following two-dimensional integral representation:
\begin{align}\bsp\label{eq:F3Def}
& F_3(\a,\a',\b,\b',\g;x,y) = 
\frac{\Gamma(\g)}{\Gamma(\b)\Gamma(\b')\Gamma(\g-\b-\b')} \times
\\
& \qquad  \int_0^1dv\int_0^{1-v}du\,
u^{\b-1}v^{\b'-1} (1-u-v)^{\g-\b-\b'-1}(1-ux)^{-\a}(1-vy)^{-\a'}\,.
\esp\end{align}
Thus we take
\bean
\omega &=& \Phi\cdot u^{n_a}v^{n_b}(1-ux)^{n_c}(1-vy)^{n_d}(1-u-v)^{n_g}\,du\wedge dv\,. \\
\Phi &=&  u^{a\ep}v^{b\ep}(1-ux)^{c\ep}(1-vy)^{d\ep}(1-u-v)^{g\ep}\,.
\eean
The five factors $P_I(u,v)$ in $\omega$ are all linear, so the geometry underlying the Appell $F_3$ function is then determined
by an arrangement of  hyperplanes corresponding to $P_I(u,v)=0$:
\begin{align}\bsp
H_a = &\{u=0\},\quad
H_b = \{v=0\},\quad
H_c = \{1-xu=0\},\\
&H_d = \{1-yv=0\},\quad
H_g = \{1-u-v=0\},
\esp\end{align}
which we represent in Fig.~\ref{fig:f3} for $x>y>1$. The dimension
of the (co)homology groups can be determined by counting the critical
points of $\Phi$. 
Since the geometry at hand is an arrangement of hyperplanes that intersect only pairwise, there is also a natural basis of homology which is the set of 
 {\em bounded chambers} (the finite connected regions in the complement of the hyperplanes) in
Fig.~\ref{fig:f3}. Either way, we see that the dimension is 4.

In order to get a convenient intersection matrix, we choose a basis of integration contours that are not exactly the bounded chambers mentioned above, but rather  four independent regions that are all triangles. Denoting them by subscripts indicating the sides of the triangles, we select
$\gamma_{abg}$ which is the original integration contour $\gamma$ used in the definition (\ref{eq:F3Def}),  $\gamma_{bcg}, \gamma_{cdg},$ and  $\gamma_{adg}$. 
In two dimensions, a canonical form for a triangle bounded by the three lines $P_1=0, P_2=0, P_3=0$ can be constructed with the following formula \cite{Matsumoto1998}:
\bean
d\log\frac{P_1}{P_2} \wedge d\log\frac{P_2}{P_3}\,.
\eean

For a basis of integrands, let us take
\bean
\varphi_{ab} &=& d\log u \wedge d\log v\,, \\
\varphi_{bc} &=& d\log(1-ux)\wedge d\log v\,, \\
\varphi_{cd} &=& d\log(1-ux)\wedge d\log(1-vy)\,, \\
\varphi_{da} &=& d\log u \wedge d\log(1-vy) \,.
\eean
Again, the notation indicates the location of the dlog singularities. This pair of bases is convenient because it leads to a diagonal intersection matrix. Since we are concerned with a nondegenerate arrangement of hyperplanes, i.e. the exponents and $x$ and $y$ are generic, and there are only normal crossings of divisors, one can read off the intersection numbers from the oriented intersections of the hyperplanes (and verify that this result agrees with the residue formula). Moreover, the intersection numbers also agree with the $\ord(\eps^{-2})$ terms of the integrals $\int_{\gamma}\Phi\varphi$: all nonvanishing terms arise from overlapping endpoint singularities.
Explicitly, the intersection matrix is given by 
 \bean
\begin{array}{c|cccc} 
{\vev{\Omega(\gamma_i),\varphi_j}}
 & \varphi_{ab} &  \varphi_{bc} & \varphi_{cd} & \varphi_{ad}  \\
  \hline
 \gamma_{abg}
 & \frac{1}{a b \eps^2} 
& 0 
& 0 
&0
 \\
\gamma_{bcg}
& 0
& \frac{1}{bc \eps^2}  
& 0 
& 0 
\\
\gamma_{cdg}
& 0
& 0 
& \frac{1}{cd \eps^2}  
& 0 
\\
\gamma_{adg}
& 0
& 0 
& 0 
& \frac{1}{ad \eps^2}  
\\
\end{array}
\eean

We thus arrive at the following coaction formula:
\bean
\Delta\left(\int_{\gamma}\Phi\varphi\right) &=&
ab\ep^2 \int_{\gamma}\Phi\varphi_{ab} \otimes \int_{\gamma_{abg}}\Phi\varphi
+ 
bc\ep^2 \int_{\gamma}\Phi\varphi_{bc} \otimes \int_{\gamma_{bcg}}\Phi\varphi 
\\ &&
 + 
cd\ep^2 \int_{\gamma}\Phi\varphi_{cd} \otimes \int_{\gamma_{cdg}}\Phi\varphi 
+ 
ad\ep^2 \int_{\gamma}\Phi\varphi_{ad} \otimes \int_{\gamma_{adg}}\Phi\varphi.
\eean
We have explicitly checked that this coaction is compatible with the coaction on MPLs in the Laurent expansion for all entries of the period matrix $\int_{\gamma_i} \Phi\varphi_j$, through weight 4.

\section{Summary and discussion}

Based on the description of generalized hypergeometric functions in terms of their integral representations and considerations of twisted (co)homology, we have proposed an algebraic coaction
on certain classes of these functions, including ${}_{p+1}F_p$, Appell $F_1, F_2, F_3, F_4$, the Lauricella series  $F_A$, $F_B$,  $F_D$, and any hypergeometric functions with a basis of integration contours consisting of positive geometries.
When the exponents in the integral representation are expanded around integer values, we claim that this coaction is \hb{compatible with the coaction on the MPLs} in the Laurent expansion. We have checked this claim to several orders in $\eps$ for ${}_{2}F_1$, ${}_{3}F_2$, and the Appell functions $F_1, F_2, F_3, F_4$. These examples, and the underlying theory, are discussed in more detail in ref. \cite{Abreu:2019wzk}.

This coaction supports our examples of \hb{Feynman-diagrammatic coaction} to all orders in the dimensional regularization parameter $\eps$.
We note that 
the hypergeometric functions appearing in known Feynman integrals actually violate a key assumption in the results related to twisted (co)homology and intersection numbers, namely that the exponents in the integral representations are nonzero and independent. However, we find that we are able to derive valid coaction formulas in degenerate limits, and we believe that these limits can be justified by a detailed analysis of twisted cycles.

A mathematical treatment of these ideas, applied to motivic versions of the Lauricella functions $F_D^{(n)}$, which includes ${}_2F_1$ and Appell $F_1$, has recently been initiated in \cite{brown2019lauricella}. The motivic version does not make use of intersection numbers, since the first and second entries in the coaction are normalized independently. The coaction argument and the first entry are motivic, while the second entry is single-valued. It would be interesting to study the connection between that treatment and ours in detail. 

Although it has not been emphasized in this brief article, the second entries in our coactions are also modified versions of the hypergeometric functions. The reason is that we require consistency with the Laurent expansion in terms of MPLs, but the second entries in the coaction of MPLs are equivalence classes modulo $2\pi i$. Therefore the closed-form hypergeometric functions in the second entry must carry a similar loss of information. A further, minor, point about the second entries of the coaction (\ref{eq:masterformula}) is that while they fit our definition of generalized hypergeometric functions, it is not always obvious how to recognize them in terms of known functions when represented by integrals over different regions. However, it is true that the second entries in the coaction belong to the same class of function, as stated in the example of eq. (\ref{eq:t1m1hyp}). In the examples we have considered here, namely ${}_2F_1$ and Appell $F_3$, it is straightforward to introduce a change of variables for each of our chosen elements of the basis of integration contours, such that the integral can be recognized as another ${}_2F_1$ or $F_3$ function, respectively.

In these proceedings we have looked only at cases with linear polynomials $P_I({\bf u})$. Nonlinear polynomials appear in examples such as ${}_{p+1}F_p$ for $p>1$, and Appell $F_4$. These functions have been considered in \cite{Abreu:2019wzk}. Since they are associated to positive geometries, we have been able to construct coactions for them as well.

\acknowledgments
This work is supported by the ``Fonds National de la Recherche Scientifique'' (FNRS), Belgium (SA),  by the 
European Research Council under the European Union's Horizon 2020 research and innovation programme through grants 
647356 (RB) and  637019 (CD), and the STFC Consolidated Grant ``Particle Physics at the Higgs Centre'' (EG, JM).

\bibliographystyle{JHEP}
\bibliography{main.bib}

\end{document}